\documentclass[conference]{IEEEtran}
\IEEEoverridecommandlockouts
\usepackage{cite}
\usepackage{amsmath,amssymb,amsfonts}
\usepackage{algorithmic}
\usepackage{graphicx}
\usepackage{textcomp}
\usepackage{siunitx}
\usepackage{lscape}
\usepackage{afterpage}
\usepackage[table]{xcolor}
\usepackage{tabularx}
\usepackage{rotating}
\usepackage{multicol}
\usepackage{subcaption}

\def\BibTeX{{\rm B\kern-.05em{\sc i\kern-.025em b}\kern-.08em
    T\kern-.1667em\lower.7ex\hbox{E}\kern-.125emX}}
\begin{document}

\title{Protecting GNSS-based Services using Time Offset Validation
}

\author{\IEEEauthorblockN{Kewei Zhang}
\IEEEauthorblockA{\textit{Networked Systems Security Group} \\
\textit{KTH Royal Institute of Technology}\\
Stockholm, Sweden \\
kewei@kth.se}
\and
\IEEEauthorblockN{Marco Spanghero}
\IEEEauthorblockA{\textit{Networked Systems Security Group} \\
	\textit{KTH Royal Institute of Technology}\\
	Stockholm, Sweden \\
	marcosp@kth.se}
\and
\IEEEauthorblockN{Panagiotis Papadimitratos}
\IEEEauthorblockA{\textit{Networked Systems Security Group} \\
	\textit{KTH Royal Institute of Technology}\\
	Stockholm, Sweden \\
	papadim@kth.se}

}

\maketitle

\begin{abstract}
Global navigation satellite systems (GNSS) provide pervasive accurate positioning and timing services for a large gamut of applications, from Time based One-Time Passwords (TOPT), to power grid and cellular systems. However, there can be security concerns for the applications due to the vulnerability of GNSS. It is important to observe that GNSS receivers are components of platforms, in principle having rich connectivity to different network infrastructures. Of particular interest is the access to a variety of timing sources, as those can be used to validate GNSS-provided location and time. Therefore, we consider off-the-shelf platforms and how to detect if the GNSS receiver is attacked or not, by cross-checking the GNSS time and time from other available sources. First, we survey different technologies to analyze their availability, accuracy and trustworthiness for time synchronization. Then, we propose a validation approach for absolute and relative time. Moreover, we design a framework and experimental setup for the evaluation of the results. Attacks can be detected based on WiFi supplied time when the adversary shifts the GNSS provided time, more than \SI{23.942}{\micro\second}; with Network Time Protocol (NTP) supplied time when the adversary-induced shift is more than \SI{2.046}{\milli\second}. Consequently, the proposal significantly limits the capability of an adversary to manipulate the victim GNSS receiver.

\end{abstract}


\section{Introduction}
The recent increased use of global satellite navigation systems (GNSS), for emerging applications, such as autonomous/unmanned vehicles or intelligent transportation systems has heightened security concerns. More so, as researchers recently demonstrated an effective GPS spoofer built with a Raspberrry Pi and a Software-Defined Radio (SDR), with a cost of only \$250 \cite{zeng2018all}; or a dual-frequency spoofer built with an SDR with a cost of only \$400 \cite{curran2018}. Therefore, any applications relying on GNSS, from mainstream mobile devices to smart vehicles, ships and large, complex systems, such as smart grids and cellular networks, face a dire risk.

A significant effort to improve security against different types of attackers, such as GNSS repeaters and spoofers, has been a central focus for both industry and the research community. GNSS vulnerabilities have been investigated in several works, e.g., \cite{zeng2018all} and \cite{kuusniemi2017feasibility,zhang2019on,gibbons2013fcc,zhang2019safeguarding}, and different countermeasures have been analyzed and evaluated \cite{anderson2017chips,Akos2012,wesson2011evaluation,papadimitratos2008,PapadimitratosJa:C:2008,PapadimitratosJ:P:2012,motella2018snap,shafiee2018detection,zhang2019secure,gulgun2019statistical}. Contributions towards protecting GNSS receivers can be divided into two main categories: countermeasures on the receiver side and on the system side. On the receiver side, one approach to detect the presence of an attacker is to check the received signal strength, e.g., through received power monitoring (RPM) (\cite{marnach2013detecting,Akos2012}) and automatic gain control (AGC) monitoring \cite{bastide2003automatic}; with special purpose hardware a receiver can determine the arriving angles of the signals from different satellites \cite{montgomery2009receiver,psiaki2014gnss,mcmilin2015field}; some work compares GNSS measurements with additional positioning information, e.g., Inertial Navigation System (INS), to detect the spoofing or replaying attacks \cite{khanafseh2014gps,curran2017use,tanil2016ins}; distortions of signal correlation function \cite{pini2013detection,ali2014vestigial,jahromi2016galileo} and clock drift (\cite{marnach2013detecting,papadimitratos2008}) can also be an indication of an attacker. 

On the system side, modification of the GNSS infrastructure is needed, to add or augment features of Signal in Space (SIS), to increase the difficulty of mounting attacks. Military signals can be encrypted with secret keys that can be accessed only by authorized entities (\cite{anderson2017chips,gsa2017}). For civilian-grade signals, Galileo currently develops navigation message authentication (NMA) for Open Service (OS) Signals \cite{caparra2016novel,humphreys2013detection,wesson2012practical}; other systems use similar approaches to protect civilian signal authenticity \cite{anderson2017chips}. However, even with NMA protection, signals can be still manipulated by sophisticated replay attacks, such as distance-decreasing attacks (\cite{zhang2019on,zhang2015gnss,zhang2019safe}) and secure code estimate-replay attacks \cite{humphreys2013detection}, which can modify each pseudorange measurement separately.

It is feasible to detect the attack by checking the consistency of the GNSS position, velocity, and time (PVT) solution. The aforementioned INS-based countermeasures \cite{khanafseh2014gps,curran2017use,tanil2016ins}, necessitate INSs. Without such hardware, it is not feasible for many applications to detect abnormalities in the PVT solution obtained from the GNSS receiver. Methods to detect GNSS spoofing attacks by checking the clock bias, within a short period were proposed (\cite{papadimitratos2008,marnach2013detecting,jafarnia2013pvt}). The detection method is based on a known linear clock state model with stable clock drift. However, it was found that the receiver's clock drift becomes stable only after about 120 minutes after switching on the receiver in room temperature, because it takes about 100 minutes for the receiver temperature to become stable \cite{marnach2013detecting}. This approach is not easily applicable, less so for a receiver in cold start. 

Currently, many commercial devices/platforms with an embedded GNSS receiver have rich connectivity, by means of different technologies, notably WLAN and cellular networks. This leads to another path for time/clock information to be used as means to detect attacks. The approach is to cross-check timing information with external time sources. Therefore, we can leverage these technologies to obtain several different external time sources, to detect if the GNSS-provided time is consistent with them. Based on this, if the external timing information source is not attacked. i.e., if it can be trusted, it is possible to determine whether the received GNSS signals are legitimate or not. The effectiveness of this approach depends on the time accuracy provided by different technologies, as discussed in Section \ref{sec:related}. Although the idea to use time as a mean of verification is not new, we propose, to the best of our knowledges, a first investigation towards generalizing the comparison of different time sources to detect discrepancies, between time provided by the GNSS and external (non-GNSS) technologies. Our experimental setup, based on commercially available off-the-shelf (COTS) devices, is a general test setup to evaluate the performance with real data.

The rest of the paper is organized as follows: Sec. \ref{sec:related} analyzes the time accuracy of different technologies; then, the adversary model is presented in Sec. \ref{sec:adversary}, following by our proposed algorithm in Sec. \ref{sec:approach}; furthermore, a test setup and evaluation results are in Sec. \ref{sec:results}; finally, Sec. \ref{sec:conclusion} concludes the work.
\section{Related Technologies}
\label{sec:related}
A GNSS receiver can be just one component in a system/platform that offers many network connectivity options. These different connections can be leveraged to obtain different external time sources independent from each other. 

The Network Time Protocol (NTP) and the Precision Time Protocol (PTP) provide accurate clock synchronization over LAN/WAN networks and they are the industry standards for synchronization in computer systems (\cite{rfc201005,ieee2002standard}). Recently, several security concerns, especially man-in-the-middle attacks and denial of service attacks, were investigated \cite{rfc201006,bishop1990security,liska2016ntp,mizrahi2017security}. NTPsec is a security-hardened implementation of NTP, which aims to make the protocol deployment compliant with more stringent security, availability, and assurance requirements \cite{raymond2016ntpsec}. The accuracy of NTP is usually within tens of milliseconds over the Internet, and it can be less than 1 millisecond in LANs with ideal network conditions. However, asymmetric network conditions and routes degrade NTP accruacy to 100 milliseconds or more (\cite{rfc1305,mills2016computer}). PTP suffers from similar problems, but it provides better accuracy, from hundreds of nanoseconds to microseconds \cite{watt2015understanding}. 
In contrast to their good performance over wired links, using these protocols over mobile communication links raises a series of challenges. One of the known problems in implementing NTP over cellular networks is the change of state of the cellular radio. If the amount of traffic on the communication link is not enough to keep the radio in active state, the radio goes in idle mode. As specified in the 3GPP documentation \cite{tsgranL2-3}, when the cellular radio is forced into a idle mode, no physical uplink or downlink are allocated. The power state transition introduces significant communication latency and it degrades the performance, limiting its accuracy. Keep-Alive messages are needed to generate enough traffic for the connection to avoid idle states \cite{Haverinen}. The achievable accuracy is within tens of ms (\cite{Haverinen,Miskinis2015}) with strong constraints on the power consumption and operational modes of the cellular radio. In scenarios where power consumption is a critical limitation, this solution is hardly applicable.

Cellular networks are widely deployed, including 2G, 3G, 4G and now 5G, providing comprehensive network access coverage in cities, highways and countryside. From the development of 3G and 4G to 5G, highly accurate time synchronization has become available. Timing Advance (TA) values, used to schedule transmissions between User End (UE) and Radio Base Station (RBS), are used to synchronize UE and RBS since 2G (\cite{3gpp200305,3gpp201605}). The TA value is normally between 0 and 63, with iincrements of 3.69 microseconds, i.e., one bit period. This value also defines the best accuracy the UE can obtain through the TA values. For LTE, Release 11 of the LTE standard defines a new System Information Block (SIB), i.e., SIB16, which contains GPS time and Coordinated Universal Time (UTC), so that the UE uses them to obtain GPS and UTC time or local time \cite{3gpp2012121960}. In 5G, two proposals for UE time synchronization methods in RAN\#81 leverage a SIB-based message, i.e., SIB16, to deliver reference time information to UEs for Time Sensitive Networking (TSN) (\cite{3gpp20181817172,3gpp20181817173}). The worst case synchronization inaccuracy with a Next Generation Node B (gNB) is expected to be $\pm$250 ns for small Industrial Internet of Things (IIoT) cells (e.g., up to 10\ m radius) \cite{3gpp20181817173}. 

Cellular links are not the only option to access high precision timing signals. WiFi and other Wireless LAN-based technologies offer several solutions. In \cite{anand2009eda}, experiments show that average propagation delay using NTP over WLAN is 2.7 ms with a standard deviation of 2.39 ms. Some customized WLAN protocols (\cite{mock2000continuous,cena2015unified}) propose storing timestamps inside Beacons, so that there is no protocol overhead in establishing synchronization between Access Point (AP) and mobile stations. It is also proposed to store many older timestamps in beacons, to ensure reliability in case of beacon loss. The accuracy these synchronization algorithms achieve is around 100 $\mu s$, with a customized driver on a Windows platform. For wireless distributed systems, synchronization of internal clocks is a fundamental problem to allow communication. Protocols such as Reference Broadcast Time Synchronization (RBS), perform well in distributed scenarios~\cite{Mahmood2017}. 
In urban environments, the high density of Access Points (APs) can provide seamless WiFi beacon coverage. This large number of beacons, generated by APs to advertise their networks, can be exploited to provide a stable flow of high-precision timing information. However, as the probability of receiving several streams of beacons is significant (given the dense deployment of APs, e.g., in urban environments), the computational power needed to process all such events can become a significant bottleneck for a low end platform. To avoid this, a subset of the available beacons, based on the proximity to the AP and the target beacon emission rate can be selected.

Local clock references in state-of-the-art CPUs \cite{pasztor2002pc} can be used to provide a very stable time, based on the timestamp instruction cycle register of the CPU. Specifically, the Time Stamp Counter (TSC) (and equivalent), a 64-bit processor register, counts the number of CPU clock cycles since reset. Therefore, the TSC value maintains very high time resolution, e.g., one nanosecond for a stable \SI{1}{GHz} processor. When the TSC is used for accurate timing, the speed/frequency of the CPU needs to be controlled and kept stable. Intel allows developers to extract TSC information since the Pentium CPU \cite{intel1997using}; similarly for AMD processors \cite{amd2006bios} and ARM processors \cite{arm2005dwt}. Performance Measurement Units (PMUs) or clock registers can be read from the Linux kernel and the user space from ARM, AMD and Intel processors. Even though platform specific, these registers are common to several architectures of the same family. 

\section{System Model}
\label{sec:adversary}
\subsection{Adversary Model}
The mathematical model to obtain the PVT solution at a GNSS receiver influenced by an adversary can be written as:
\begin{equation}
\mathbf{y=Hx + f + v}
\end{equation}
where $\mathbf{y}$, $n\times 1$ vector, contains pseudorange measurements of the receiver to $n$ satellites; $\mathbf{H}$ is a $n \times 4$ observation matrix; $\mathbf{x}=[x, y, z, \delta t]$, is the receiver state, including three-dimension coordinates and clock offset; $\mathbf{f}$ is a $n\times 1$ offset vector that an adversary introduces to the pseudorange measurements; $\mathbf{v}$, $n \times 1$ vector, is noise. Generally speaking, there is no limitation on how the adversary mounts the attacks, e.g., by replaying previously recorded signals or by transmitting fine-grained simulated signals, and the adversary objectives. 

Considering a simple adversary, all the elements in $\mathbf{x}$, including $x,\ y,\ z$ and $\delta t$, are manipulated when the adversary induces a specific victim receiver location, or the adversary could seek to mislead the receiver to follow the adversary-intended time, by inducing a specific $\delta t$. Then, our and any time cross-checking proposal will detect the attack when the change in $\delta t$ exceeds a certain threshold. The objective is to design time-based validation (or attack detection) that operates in a way that severely limits the adversary. Intuitively, the approach detects the lowest discrepancy caused by an attack. A sophisticated adversary, seeking to change the victim's location without modifying the victim's time \cite{humphreys2012texas} cannot be detected by any clock-related countermeasure, and thus by our proposal either.

The GNSS receiver may be at cold start or in a state of continuously tracking satellites. In the cold start, the system needs to acquire absolute time from satellites or other external time providers. When the GNSS receiver already tracks satellites, the adversary can either first jam the signals reception at the receiver, then transmit recorded/forged signals to the receiver, or use a signal lift-off technique to take over the receiver.

\subsection{System Assumptions}
Maintaining synchronization of the GNSS receiver as accurate as possible is not the goal of our proposal. Instead, we aim at evaluating to which extent the existing external time transfer technologies can be used towards the GNSS-provided time and location verification. When the external time technologies are used to detect the manipulation of the GNSS-provided information, clear assumptions on the trustworthiness and accuracy metrics are needed: 
\begin{itemize}
	\item It is not likely that the adversary can attack the victim GNSS receiver and at the same time compromise the external time sources or manipulate the access to the external, non-GNSS, time sources, e.g., NTP servers. Therefore, for this work, we assume that external time sources are trusted. For example, the network access per se can be encrypted and authenticated, or the time-providing network server or component (e.g., access point or base station) can be authenticated.
	
	\textbf{Remark:} In the event of no trusted external non-GNSS time sources, the approach would amount to a discrepancy detection between GNSS-provided time and one or more external time sources. Intuitively, such a discrepancy detection between two essentially non-trusted sources of time can still be useful: it can reveal that either the GNSS or the external time source(s) is attacked. This more complex attack surface warrants its own investigation; we discuss this briefly in Sec. \ref{sec:conclusion}. 
	
	\item The GNSS receiver is always the primary synchronization source, unless it is deemed not trusted by the validation/detection scheme.
	\item The system always chooses the most accurate time source. The only exception: another available source more trusted than the default one, even if the latter is less accurate.
	\item The system will always synchronize with the most trusted available source, informing the user about any changes in time reference.
\end{itemize}
\section{Solution Approach}
\label{sec:approach}
Without loss of generality, we consider two situations: 1) there is only one  external time technology, e.g., due to limited connectivity or functionality; 2) multiple available external time technologies are available. Moreover, the approach can be developed in two different directions: 1) validating relative time that is, the difference of the GNSS and external technologies elapsed time during the same time interval; 2) validating absolute time, that is, the difference of absolute time from each technology (GNSS or not). For the sake of simple presentation, we discuss first the case of a single available external technology. Then, we extend it to the case of multiple available external technologies.
\subsection{Single Available External Technology}
\begin{figure}[!t]
	\centering
	\includegraphics[width=0.8\linewidth]{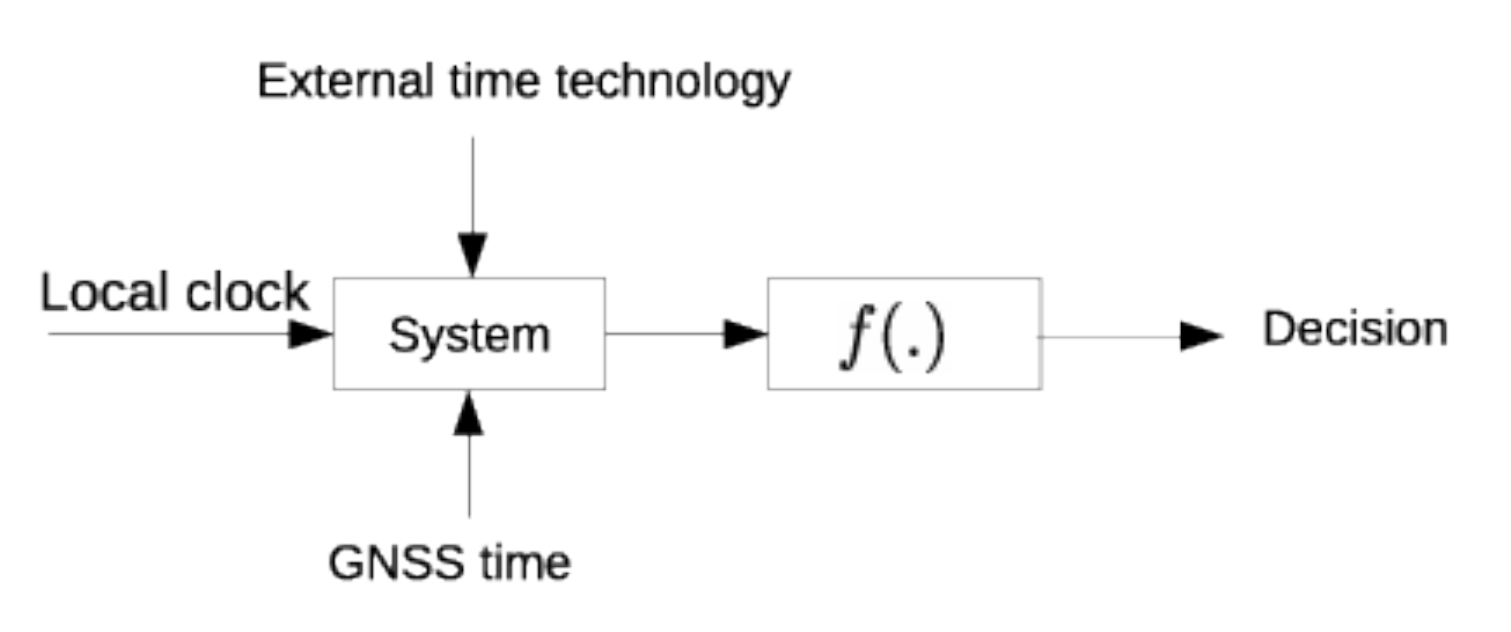}
	\caption{Illustration of the approach with a single external time source}
	\label{fig:stru1_approach}
\end{figure}

In general, the device/system can access different external time sources/technologies, e.g., WLAN or cellular networks, each providing a different level of accuracy. However, due to environment limitations and other constraints, there might be only one available external source. As Fig. \ref{fig:stru1_approach} shows, the system applies a function, $f(.)$, to the GNSS time and the one provided by the external technology. The output indicates whether the GNSS time is consistent with the external time sources.

This function, $f(.)$, can be implemented based on two approaches:
\begin{itemize}
	\item Validating absolute time, $T$
	\item Validating relative time, $\Delta t=T_n-T_{n-1}$, where n is the index of GNSS time update
\end{itemize}

\subsubsection{Absolute time checking}
\begin{equation}
f(t)=f(|T_{ext}(t)-T_{GNSS}(t)|)
\label{eq:f1}
\end{equation}
is a function of the time difference between GNSS and the external technology. Specifically, $T_{ext}$ is the time value from the external source, $T_{GNSS}$ is the GNSS-provided time, and $t$ is a time instance at the system when both $T_{ext}$ and $T_{GNSS}$ are available.

The receiver starts acquiring time information from the GNSS, in cold or warm start, and simultaneously acquires time from the external technology; it updates its GNSS-provided time every $\tau$ seconds, with the value, $\tau$, depending on the design of the receiver, e.g., \SI{500}{\milli\second} or \SI{1}{\second}. For each GNSS update, there is a time fetch from the external time technology. We do not consider the accuracy of GNSS-provided time, i.e., around \SI{100}{\nano\second}, which means that the GNSS-provided time is deemed accurate in the absence of an adversary. Therefore, we have:
\begin{equation}
f(t)<\epsilon_{ext}
\label{eq:ext_absolute}
\end{equation}
where $\epsilon_{ext}$ is the external time technology accuracy, subject to the network delays, the wireless propagation environment or the external time source attached master clock. 

For each available technology, there can be multiple time information sources. For instance, our enhanced receiver can acquire NTP time from several different NTP servers; or it can receive WiFi beacons from multiple access points simultaneously. Therefore, assuming there are $k$ time sources for a single technology, we have series of $f(t)$ values for each time source:

\begin{equation}
\begin{split}
& f^1(t_1) \quad f^1(t_2) \quad \dots \quad f^1(t_n) \\
& \qquad \qquad \qquad \vdots\\
& f^k(t_1) \quad f^k(t_2) \quad \dots \quad f^k(t_n) \\
\end{split}
\label{eq:series_data}
\end{equation}
 By assuming the $k$ sources of one technology (having same attributes and thus expected time accuracy), we set a counter \textit{m} incremented at each time instance $t_n$, if Eq. \ref{eq:ext_absolute} is true:
\begin{equation}
\text{For }  i=1,...,k; \qquad m=m+1 \quad \text{if } f^i(t_n)<\epsilon_{ext}
\label{eq:voting}
\end{equation}
Then, the approach makes a decision at time $t_n$ based on:
\begin{equation}
\frac{m}{k}>\frac{1}{2} \text{ (or a desired higher value)}
\label{eq:decision}
\end{equation}
which indicates that the majority of sources of the available time technology satisfy Eq. \ref{eq:ext_absolute}, i.e., the technology-specific accuracy threshold.

If Eq. \ref{eq:decision} is not true, this indicates a discrepancy between time acquired by GNSS and the external technology.

\subsubsection{Relative time checking}
\begin{equation}
f(t)=f(|\Delta t_{{ext}}(t)-\Delta t_{{GNSS}}(t)|)
\label{eq:f2}
\end{equation}
where $\Delta t_{{ext}}(t)=T_{ext}(t+1)-T_{ext}(t)$ and $\Delta t_{GNSS}(t)=T_{GNSS}(t+1)-T_{GNSS}(t)$. The idea of relative-time checking is that, given one interval measured by GNSS-provided time, the elapsed time measured by the external technology should be within a certain threshold. In absence of an adversary, $f(t)$ satisfies the following:
\begin{equation}
f(t)<\epsilon_{ext}
\label{eq:ext_relative}
\end{equation}

The validation process is similar as described in Eqs. \ref{eq:series_data} and \ref{eq:decision}. 

For both absolute-time and relative-time checking, in order to reduce the false alarm probability, we can extend this scheme to an aggregated scheme: the approach makes one decision every Q time instances. When any Q successive events give negative results based on Eq. \ref{eq:decision}, an attack or discrepancy of GNSS-provided time and the external technology provided time is signaled.

\subsection{Multiple External Available Technologies}
\label{sec:multiple_ext_ts}
\begin{figure}[!t]
	\centering
	\includegraphics[width=0.95\linewidth]{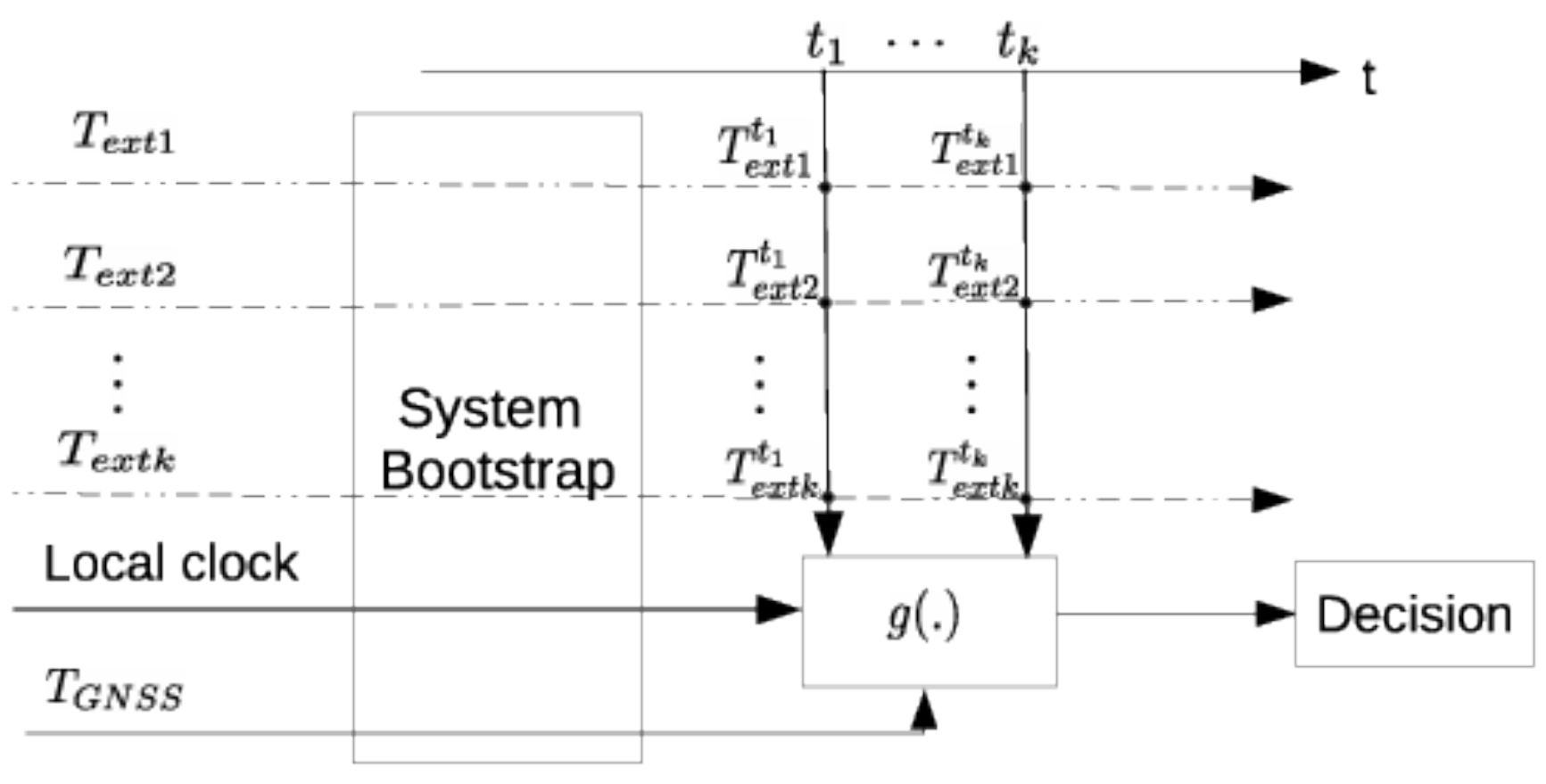}
	\caption{Illustration of the approach with multiple external time sources}
	\label{fig:stru2_approach}
\end{figure}

Multiple external technologies can be available in many off-the-shelf platforms. As Fig. \ref{fig:stru2_approach} shows, during the system bootstrapping phase, the system searches and locks to available satellites, thus obtains PVT solutions. Meanwhile, it acquires time information from other external sources. The system has a predefined setting about the accuracy (and trustworthiness, in the next version of this work, as currently external time sources are deemed trusted) of different external time technologies, according to historical statistics.

Therefore, for absolute-time and relative-time checking at time instance $t$, $g(t)$ is defined as follows:
\begin{equation}
\begin{split}
&g(t)=g\{|T_{ext1}(t)-T_{GNSS}(t)|,\dots,|T_{extk}(t)-T_{GNSS}(t)|\} \\
&g(t)=g\{|\Delta t_{{ext1}}(t)-\Delta t_{{GNSS}}(t)|,\dots, \\
&\qquad \qquad |\Delta t_{{extk}}(t)-\Delta t_{{GNSS}}(t)|\}
\end{split}
\label{eq:multiple}
\end{equation}

When we trust all the external non-GNSS time sources (of different types/technologies), if the majority of those time technologies fulfills Eq. \ref{eq:decision}, the approach deems that GNSS provided time is not faulty.

If we do not fully trust the external time technologies, the approach applies a weight to each time technology, by considering their level of trustworthiness and accuracy. For instance, in an industrial environment, the trusted WLAN access points have higher weight than cellular networks. Hence, the decision at each time instance is made based on:
\begin{equation}
\begin{split}
g(t)&= w_{ext1}\frac{f_{ext1}(t)}{\epsilon_{ext1}}+w_{ext2}\frac{f_{ext2}(t)}{\epsilon_{ext2}} +\dots+w_{extk}\frac{f_{extk}(t) }{\epsilon_{extk}} \\
&<1
\end{split}
\label{eq:gt}
\end{equation}
where $w_{extk}$ is the weight of $k^{th}$ time technology and $\sum_{i=1}^{k}w_{exti}=1$ for $k$ external time technologies, and $f(t)$ can be a function based on absolute time, Eq. \ref{eq:f1}, or relative time, Eq. \ref{eq:f2}. The weights, $w_{extk}$, can be defined based on the trustworthiness and accuracy of external time sources and the conservativeness of threshold $\epsilon_{extk}$. In order to reduce the false alarm probability, aggregation of results of successive $Q$ decisions, to obtain one final decision on the attack, can be used.

\section{Architecture and Evaluation}
\label{sec:results}

\subsection{Framework/Architecture}
\begin{figure*}[!t]
	\centering
	\begin{subfigure}{0.49\textwidth}
		\centering
		\includegraphics[height=2.5in]{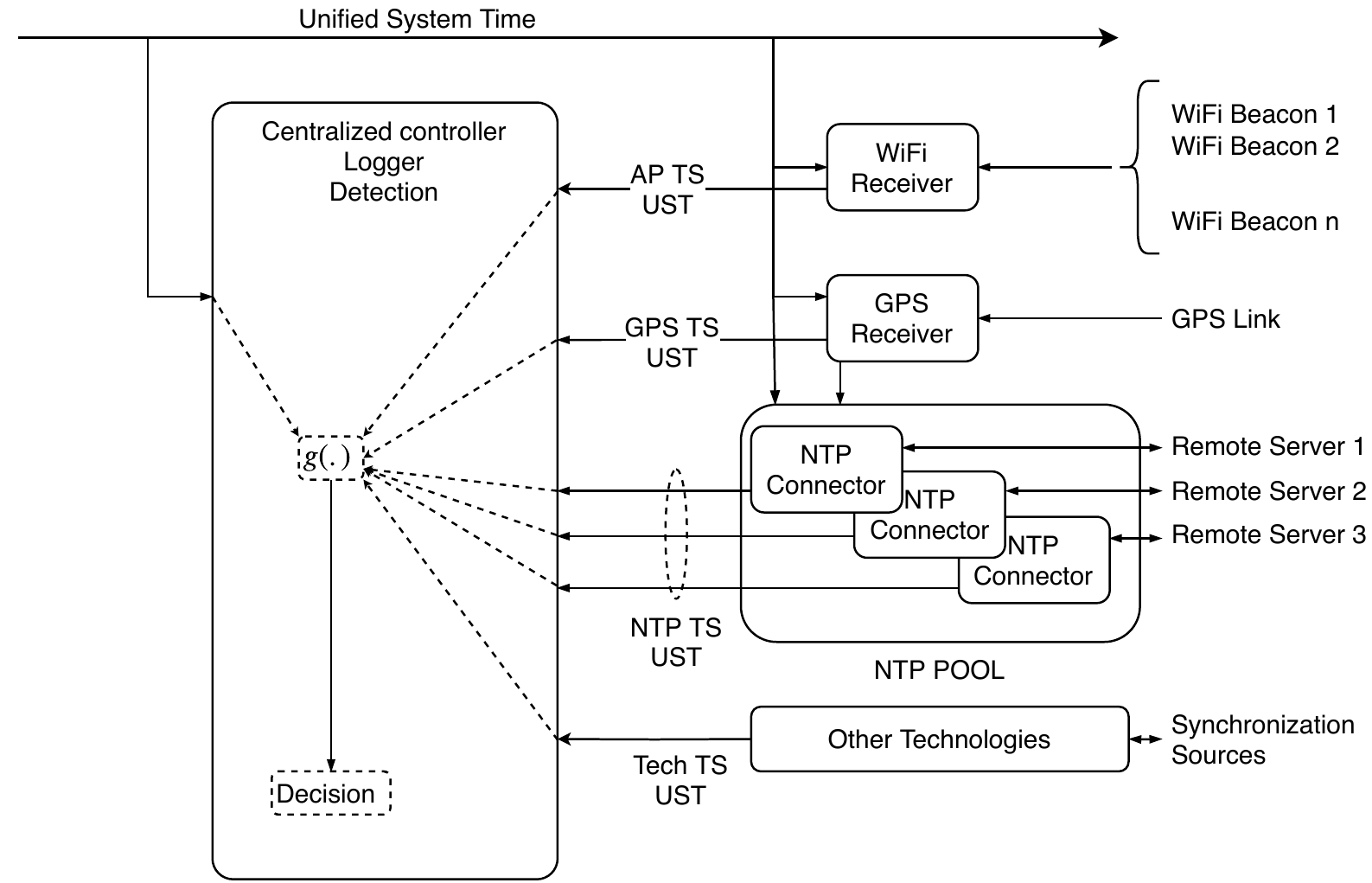}
		\caption{Framework architecture}
		\label{fig:system}
	\end{subfigure}
	~
	\begin{subfigure}{0.49\textwidth}
		\centering
		\includegraphics[height=2.5in]{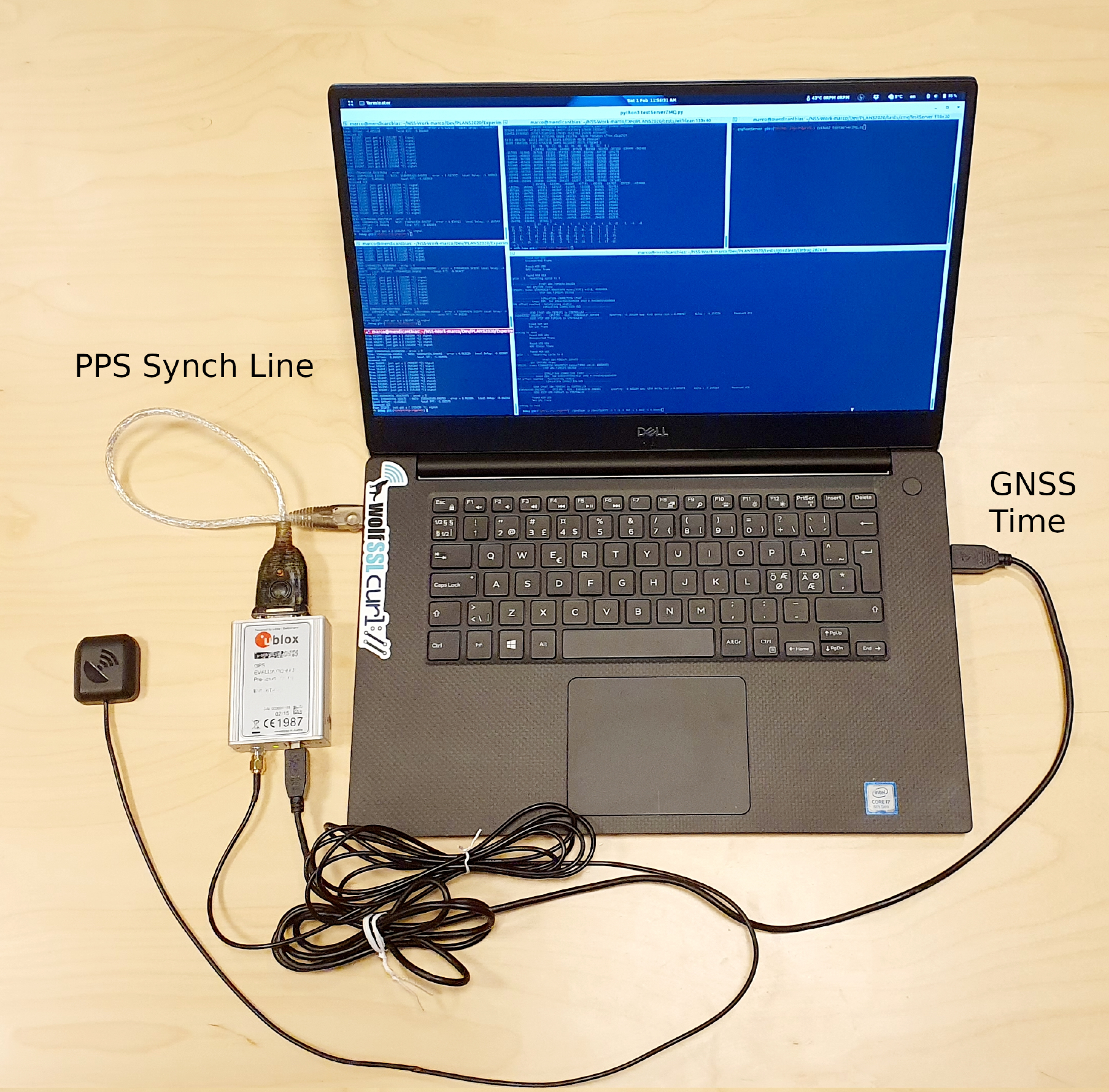}
		\caption{Measurement and evaluation setup}
		\label{fig:setup}
	\end{subfigure}
	\caption{Framework architecture and demonstration setup}
\end{figure*}

To evaluate the concept and demonstrate the results, we designed an architecture compatible with multiple external time technologies, as presented in Fig. \ref{fig:system}. We have a centralized controller that interacts with the system clock and all other time technologies, including GPS, WiFi beacons, NTP servers, etc. The system detection is triggered by the GPS time update within a specified interval; the system makes a detection decision at each specified interval with the available collected data. 

One of the challenges is the combination of synchronous and asynchronous data collection processes from different time sources. The reason is that when the system triggers a detection event, both for absolute time verification and relative time verification, all the collected data must be aligned, in order to compare them with each other. More specifically, data collection of WiFi beacons is asynchronous due to their spontaneous transmission characteristics; NTP data collection is synchronous because the NTP request is on-demand when the system attempts a request. 

The solution is to apply a time alignment for different time technologies, especially for the asynchronous data collection. We use the platform/system clock as a reference to align the data; the system timestamps each data collection with its local clock and compensates for delays of data provided by different time technologies.

\subsection{Experimental Setup}
We leverage two external time services, NTP and WiFi beacons, to verify the GPS time, as presented in Fig. \ref{fig:setup}. The ublox EVK-6T evaluation kit \cite{ublox2012} offers two interfaces for data transmission: a USB2 port provides a real-time PVT solution that the manipulated GPS time is synthesized based on; a RS232 serial port provides a GPS time pulse for the host synchronization. The serial port provides a high accuracy Pulse Per Second (PPS) via the Data Carrier Detect (DCD) pin, which is used as a reference to check the performance of the evaluation results.

Beyond the GPS receiver, the rest of the configuration includes:
\begin{itemize}
	\item Host machine: Intel I7 CPU running a Linux system whose kernel supports high precision timing.
	\item Host WiFi card: Intel Corporation Wireless-AC 9260.
	\item WiFi beacons: from surrounding access points of the office building.
	\item NTP servers: three servers in Sweden.
	\item GPS PVT rate: \SI{1}{Hz}.
	\item Observation window candidates for APs: $T_{window}=\{1024,\ 3072\ ,5120\}$~ms.
\end{itemize}

\subsection{Evaluation}

\subsubsection{Accuracy Analysis}
To validate the GPS time, we need to obtain the accuracy, $\epsilon_{ext}$, of each technology, as shown in Eqs. \ref{eq:ext_absolute} and \ref{eq:ext_relative}.
The accuracy is obtained by comparing the time information from each technology with a GNSS disciplined oscillator. 

\begin{figure}[!t]
	\centering
	\includegraphics[width=\linewidth]{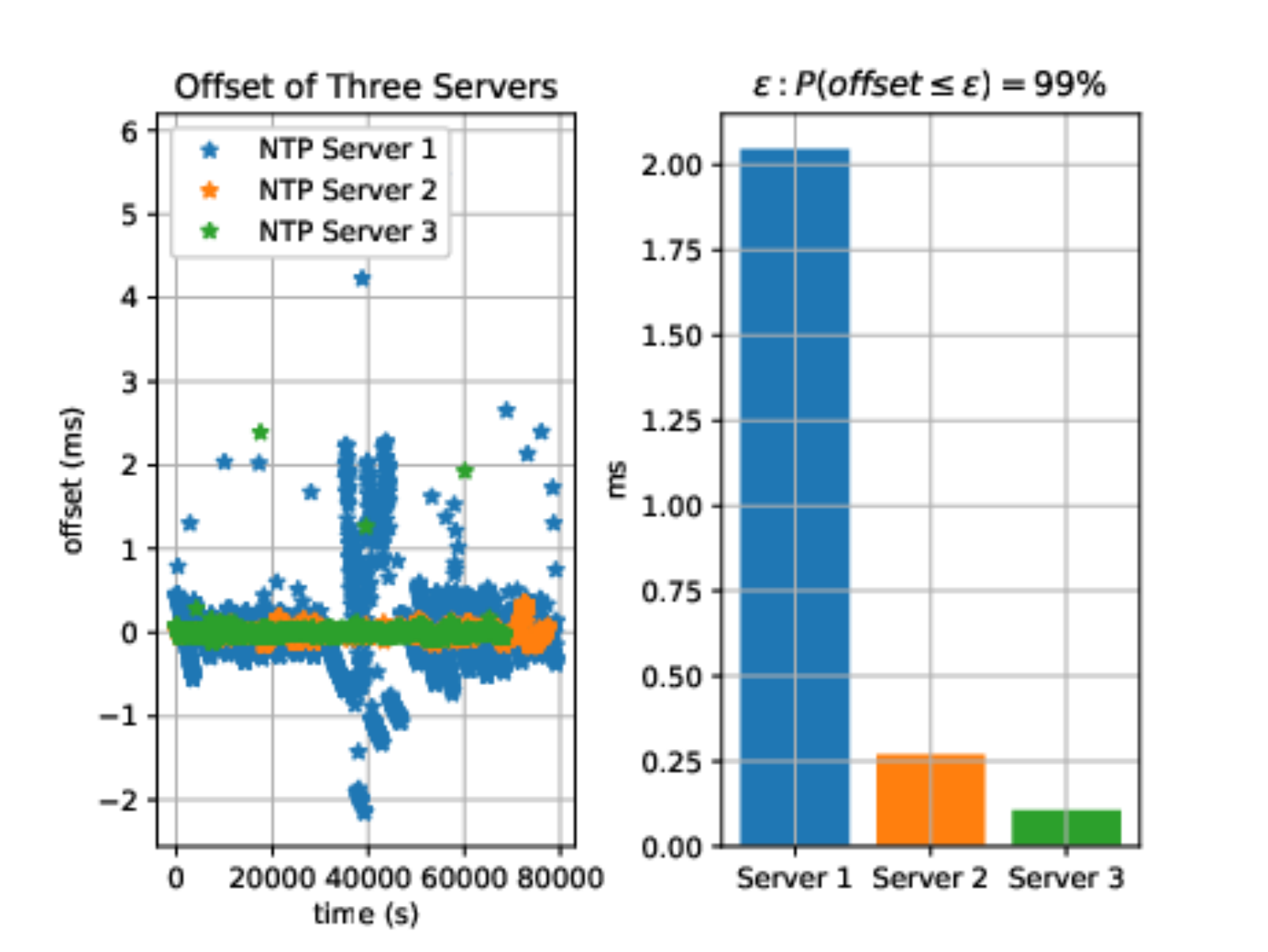}
	\caption{Statistics of three NTP servers}
	\label{fig:ntp_accuracy}
\end{figure}

For the case of NTP, the accuracy is calculated based on the offset between the GNSS-provided time and the NTP server provided time. The left plot of Fig. \ref{fig:ntp_accuracy} shows the offset of three different NTP servers located in the same country, for a period of 22 hours. We use the 99\% quantile of each server offset as its accuracy, as shown in the right plot of Fig. \ref{fig:ntp_accuracy}, to represent our parameter $\epsilon_{ext}$. The highest value among the three $\epsilon_{ext}$ is chosen to set the threshold for the NTP time. Therefore, we have:
\begin{equation}
\epsilon_{NTP} = 2.046 \text{ ms}
\label{eq:ntp_accuracy}
\end{equation}

\begin{figure*}[!t]
	\centering
	\includegraphics[width=0.9\textwidth,height=2.9in]{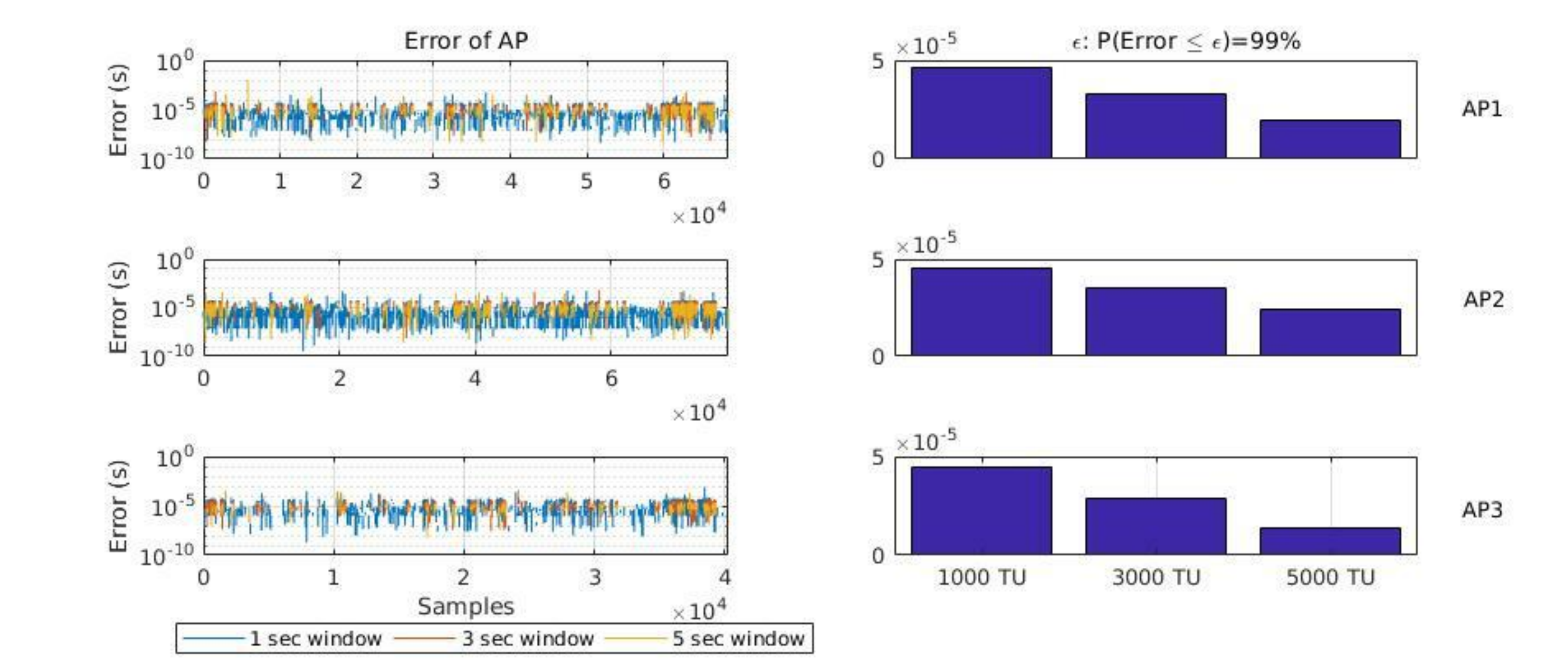}
	\caption{Statistics of three access points (AP)}
	\label{fig:ap_accuracy}
\end{figure*}
A similar approach is used to profile WiFi beacons. By default, APs transmit beacons at a 100 Time Unit (TU) interval which corresponds to 1024 microseconds \cite{geier2002802}. In the IEEE 802.11 standard, a timestamp field contained in each beacon indicates the time, notably $T_{AP}^B$, the beacon leaves the AP. Specifically, $T_{AP}^B$ is the elapsed time since power-up of the radio interface. The accuracy of the WiFi beacons is calculated over three different observation windows, $T_{window}=\{1000,\ 3000,\ 5000\}$~TU. First, the difference between the timestamp of the received beacon at the end of the observation window and the timestamp of the beacon at the beginning of the window is $T^{B_{end}}_{AP}-T^{B_{begin}}_{AP}$. Then, the error between this difference and the $T_{window}$ interval measured by the GNSS-disciplined clock is used to determine the accuracy for WiFi beacons, as presented in Fig. \ref{fig:ap_accuracy}. The left plot is the error of recorded AP data during each $T_{window}$ compared to the one measured by the GNSS disciplined clock, and the right plot gives their 99\% quantiles.

With a similar approach to the one used to evaluate the NTP accuracy, we choose the highest value of the quantiles among the three APs for each $T_{window}$ as the accuracies:
\begin{equation}
\epsilon_{WiFi} = \Big\{ \begin{array}{cc}
46.064 \text{ $\mu$s}& \qquad T_{window}=1024 \text{ ms} \\
35.021 \text{ $\mu$s}& \qquad T_{window}=3072 \text{ ms} \\
23.942 \text{ $\mu$s}& \qquad T_{window}=5120 \text{ ms}
\end{array}
\label{eq:wifi_accuracy} 
\end{equation}

\subsubsection{Evaluation Results}
\begin{figure}[!t]
	\centering
	\includegraphics[width=0.95\linewidth]{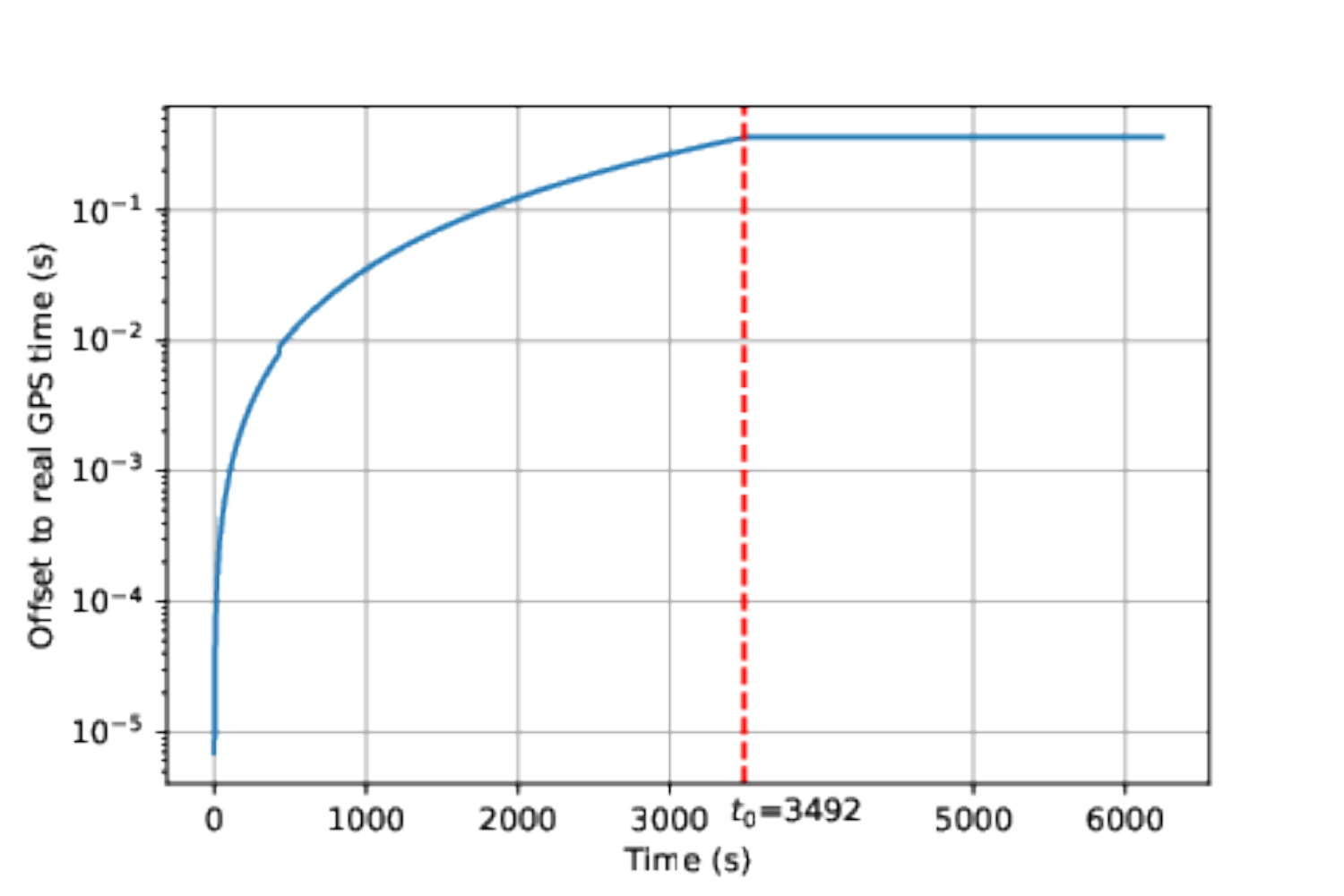}
	\caption{The offset between real GPS time and synthesized GPS time}
	\label{fig:gps_offset}
\end{figure}

The synthesized GPS time we use is obtained by applying an offset function, $of(.)$, to the real GPS time at step $n$:
\begin{equation}
\begin{array}{l}
of(1) =t_{bias} \\
of(2) = of(1) + \beta \\
\qquad \vdots \\
of(n-1) = of(n-2)+(n-3)*\beta \\
of(n) = of(n-1)+(n-2)*\beta\\
\end{array}
\end{equation}
where $t_{bias}$ is the initial offset to the real GPS time, $\beta$ controls the rate of increment of the offset and $n$ indicates the $n^{th}$ GPS update. Fig. \ref{fig:gps_offset} shows our synthesized GPS attack, $t_{bias}=$\SI{5}{\micro\second} and $\beta=$\SI{0.055}{\micro\second}. After time $t_0=3492$, the offset is maintained constantly, i.e., \SI{360}{\milli\second}.

For the evaluation based on relative time checking with WiFi beacons, for the three different observation windows, the system selects the GPS time samples every $\{1,\ 3,\ 5\}$ updates, then it picks beacons within the window as follows:
\begin{equation}
\begin{array}{l}
\text{start beacon $B_1$: }|T_{RX}^{B_1} - T_{window}^{start}|<\SI{100}{TU}\\
\text{end beacon $B_2$: }|T_{RX}^{B_2} - T_{window}^{end}|<\SI{100}{TU} \\
\end{array}
\end{equation}
where $T_{RX}^{B}$ is the system reception time at the receiver, and $T_{window}^{start}$ and $T_{window}^{end}$ are the system instants of GPS updates at the beginning and end of the window.

\begin{figure*}[!t]
	\centering
	\begin{subfigure}{0.32\textwidth}
		\includegraphics[width=\linewidth]{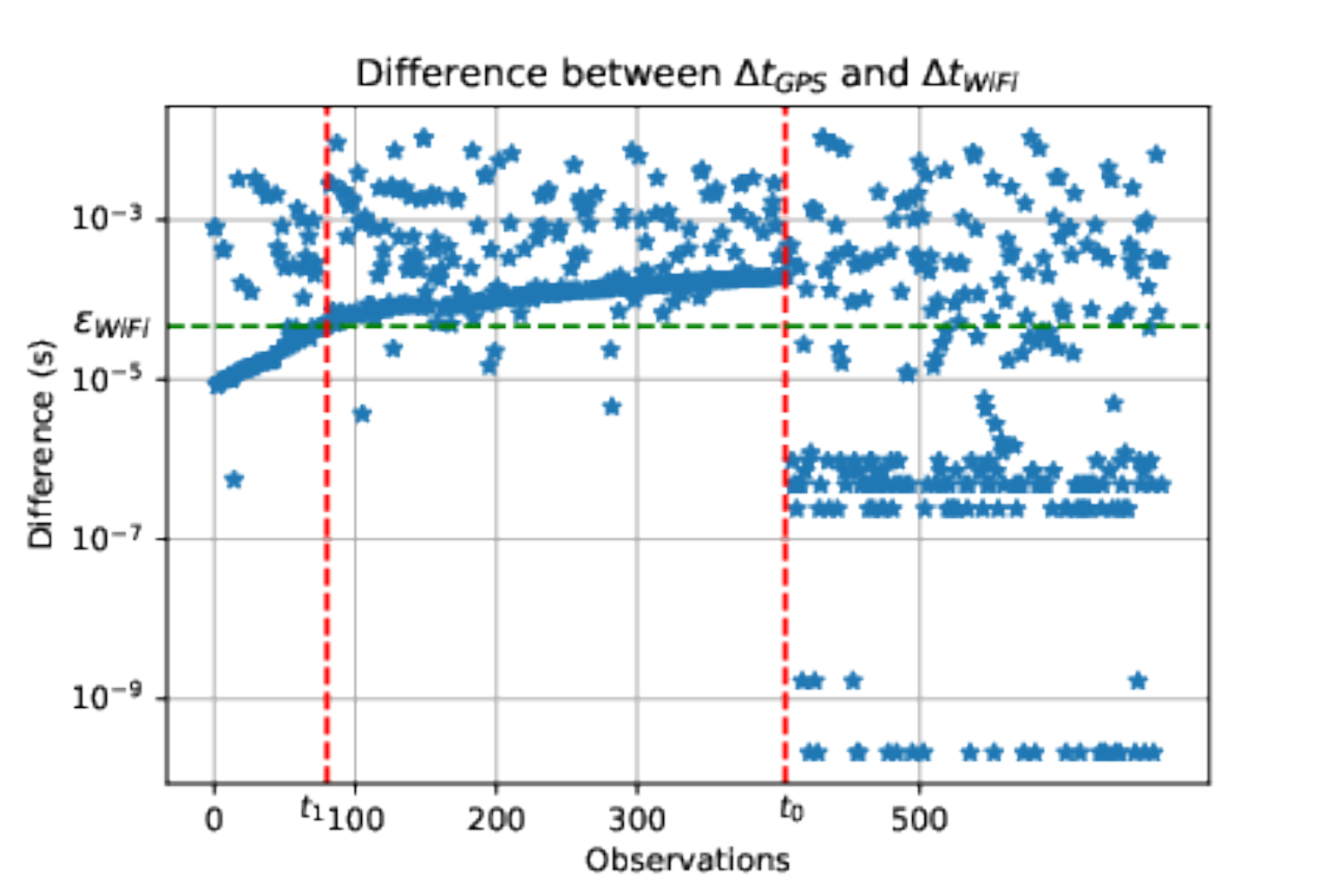}
		\caption{$T_{window}=$ \SI{1024}{\milli\second}}
		\label{fig:wifi_results1}
	\end{subfigure}
	\begin{subfigure}{0.32\textwidth}
		\includegraphics[width=\linewidth]{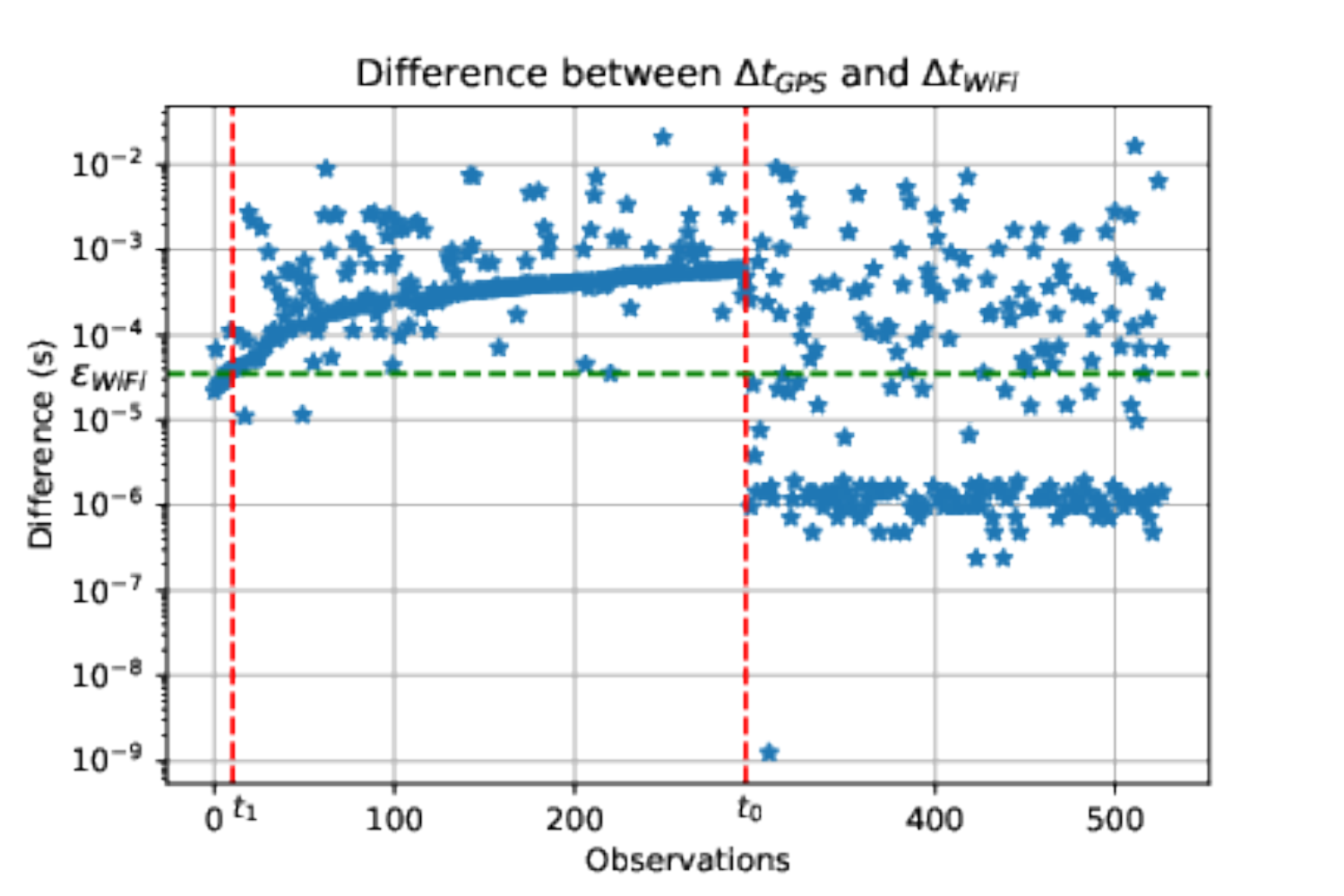}
		\caption{$T_{window}=$ \SI{3072}{\milli\second}}
		\label{fig:wifi_results3}
	\end{subfigure}
	\begin{subfigure}{0.32\textwidth}
		\includegraphics[width=\linewidth]{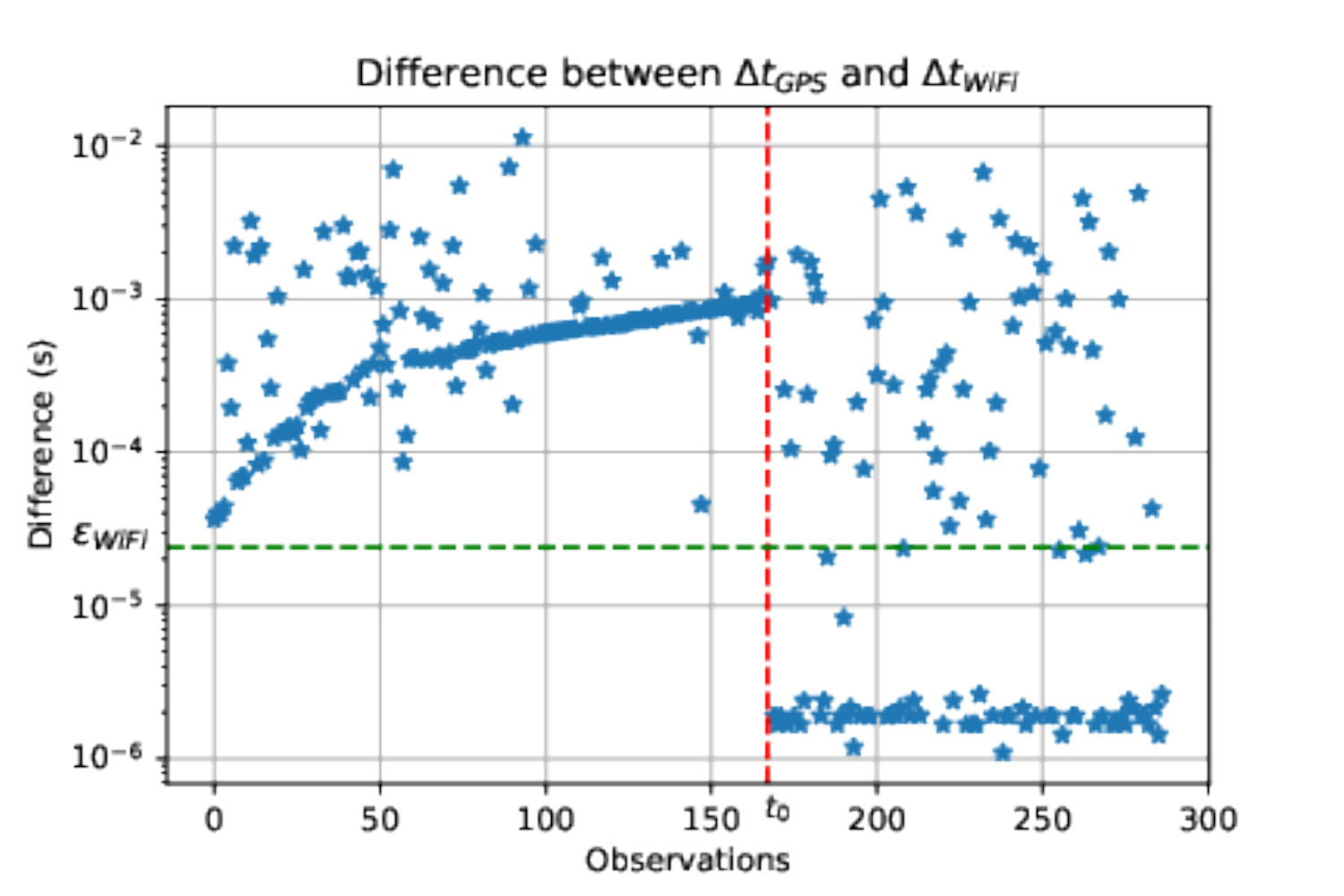}
		\caption{$T_{window}=$ \SI{5120}{\milli\second}}
		\label{fig:wifi_results5}
	\end{subfigure}
	\caption{Verification results of WiFi beacons for different $T_{window}$}
	\label{fig:verification_wifi}
\end{figure*}

When beacons satisfying the above requirements cannot be found, the system triggers a detection at the next GPS time update. Otherwise, it applies $T_{AP}^{B_2}-T_{AP}^{B_1}$ as $\Delta t_{{ext}}$ to Eq. \ref{eq:f2} and tests the comparison against the accuracy threshold in Eq. \ref{eq:wifi_accuracy}. The test results are presented in Fig. \ref{fig:verification_wifi}; where we can see the scheme cannot detect the GPS attack in the beginning, before time $t_1$, when the offset is lower than $\epsilon_{WiFi}$ for $T_{window}= \{1000,\ 3000\}$ TU. The scheme can detect the attack from time $t_1$ to $t_0$. But it can trigger an alarm of detecting the attack for $T_{window}= \{5000\}$ TU from the beginning of the attack to $t_0$. After time $t_0$, because the time offset is constant, the relative time checking solution with WiFi beacons is no longer effective.

\begin{figure*}[!t]
	\centering
	\begin{subfigure}{0.32\textwidth}
		\includegraphics[width=\linewidth]{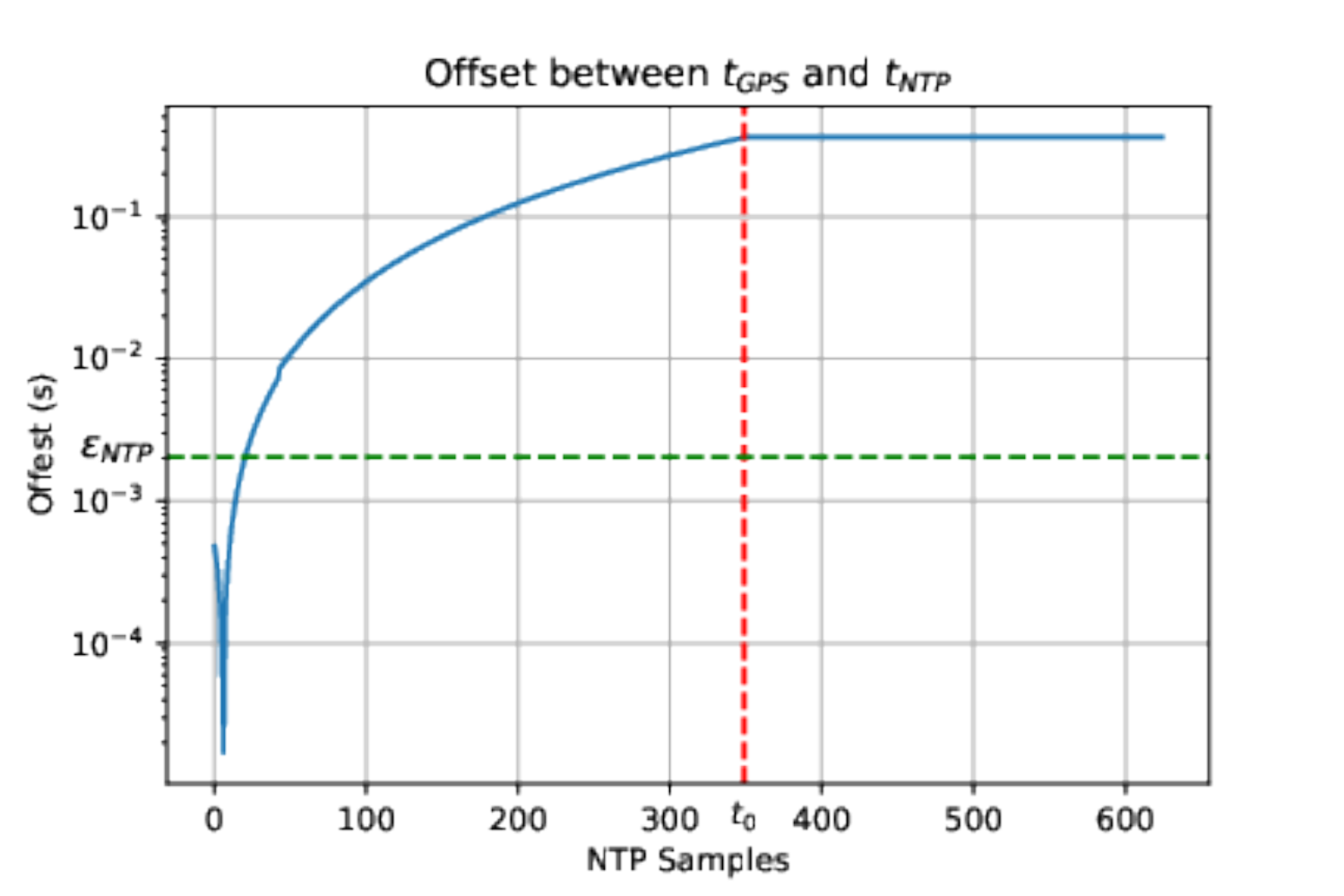}
		\caption{Server 1}
		\label{fig:ntp_results0}
	\end{subfigure}
	\begin{subfigure}{0.32\textwidth}
		\includegraphics[width=\linewidth]{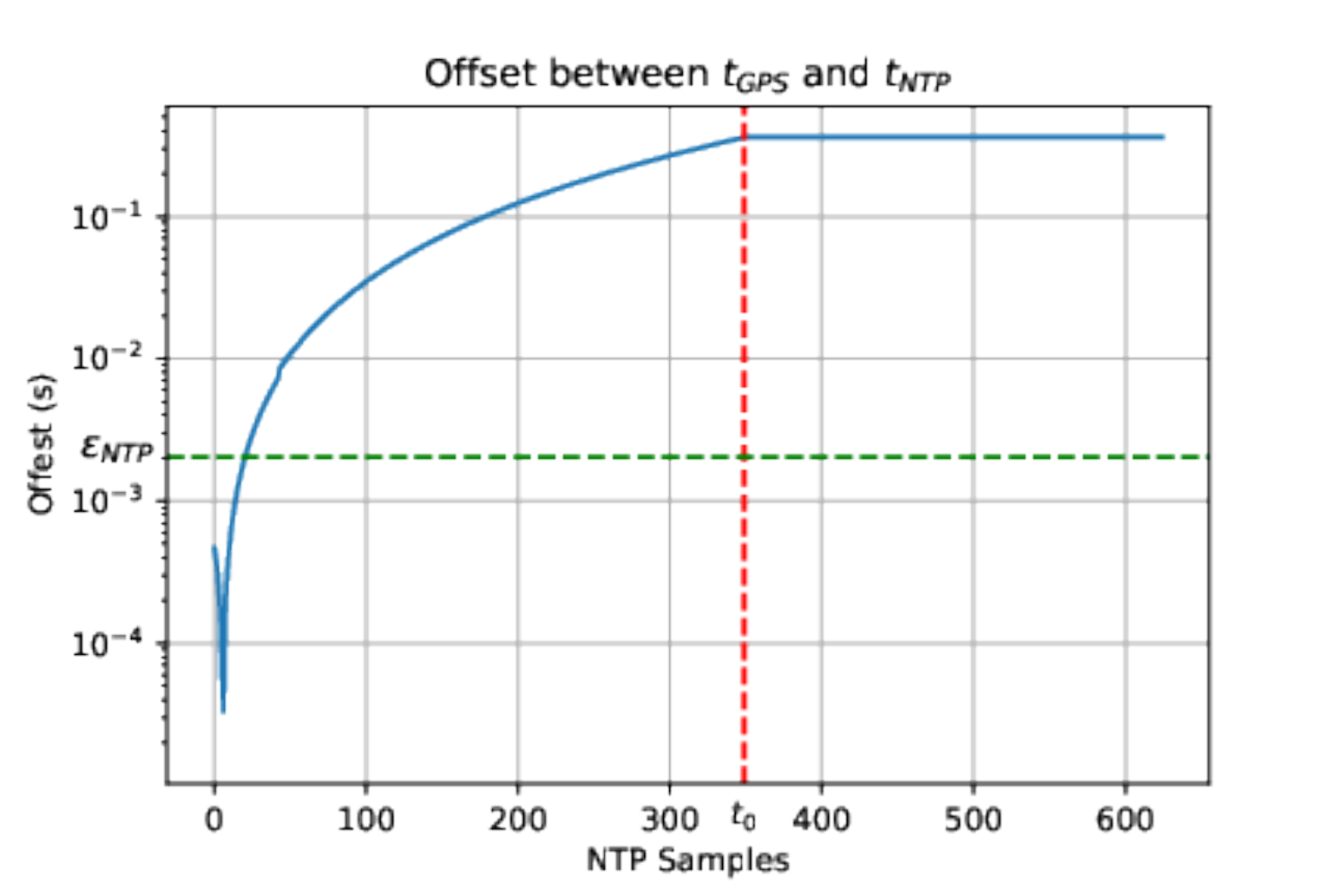}
		\caption{Server 2}
		\label{fig:ntp_results1}
	\end{subfigure}
	\begin{subfigure}{0.32\textwidth}
		\includegraphics[width=\linewidth]{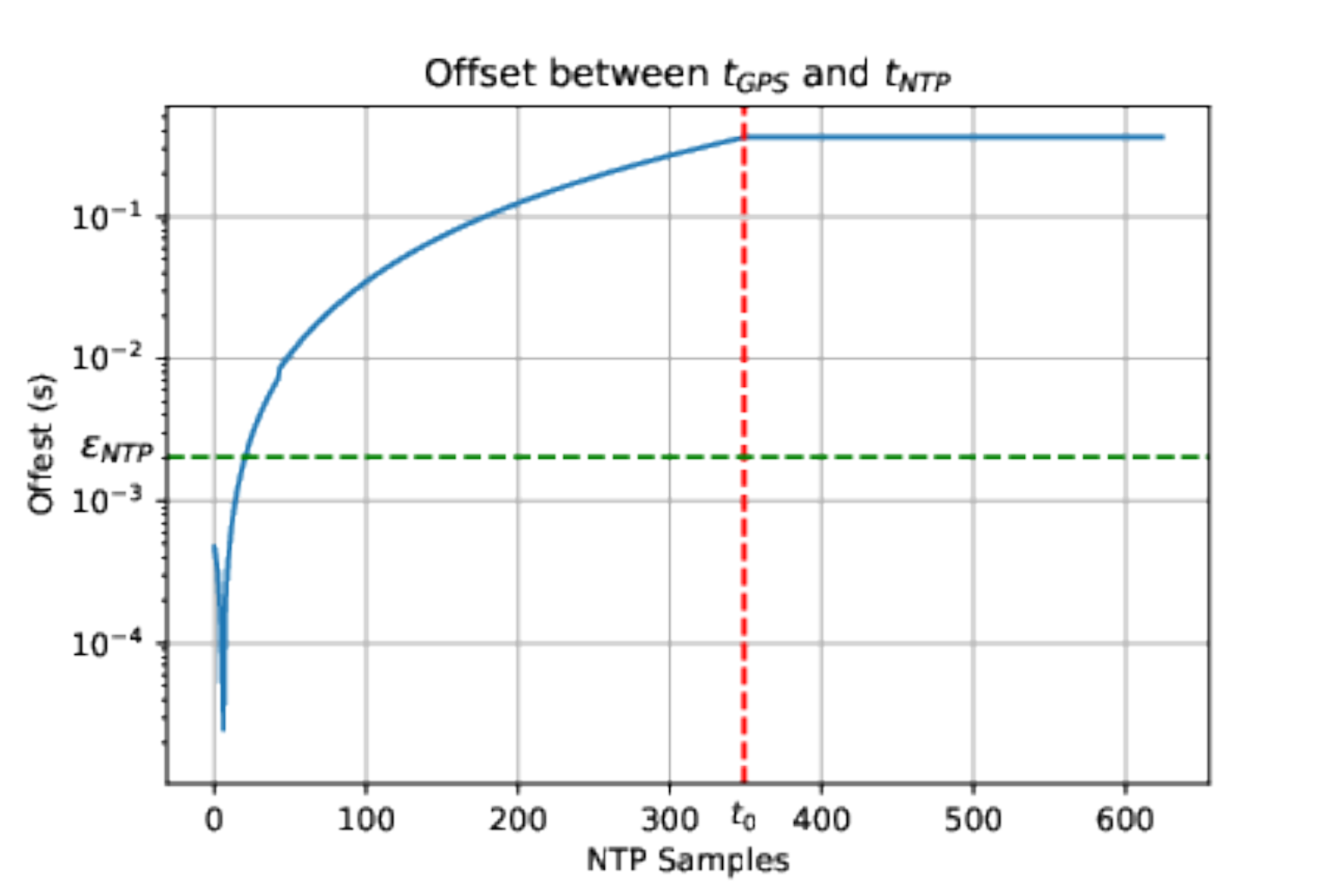}
		\caption{Server 3}
		\label{fig:ntp_results2}
	\end{subfigure}
	\caption{Verification results for different NTP servers}
	\label{fig:verification_ntp}
\end{figure*}

For the verification based on absolute time checking with NTP, the system acquires the NTP values at selected GPS samples, with a frequency lower than the minimum polling frequency specified by the NTP service provider. The system applies the acquired $T_{ext}$ and corresponding $T_{GPS}$ to Eq. \ref{eq:ext_absolute}, with the accuracy defined in Eq. \ref{eq:ntp_accuracy}. The comparison results for each NTP server are presented in Fig. \ref{fig:verification_ntp}: the system can detect the attack when the offset is higher than $\epsilon_{NTP}$. 

\section{Discussion and future work}
\label{sec:conclusion}
We investigated to which extent other timing technologies can potentially protect a GNSS receiver from a replay/spoofing attack, based on their availability, diffusion and accuracy. We proposed a scheme and designed an experimental framework for the detection of these attacks. The scheme and the framework can integrate various timing technologies, providing a scalable and flexible solution based on time cross-checking. We demonstrated and evaluated the proposal with a real-world collected data, with different setup configurations, both for relative time checking and absolute time checking. 

The proposed concept aims to protect a GNSS receiver using time information obtained from external sources and alternative independent technologies. One of the concerns that can arise is to which extent we can trust such external information. It is part of future investigation to determine how the level of trustworthiness of these technologies affects the security of the proposal. Without authentication of the WiFi beacons or authenticated network access or authenticated time servers, a substantial limitation would arise. Broadcasting bogus WiFi beacons is not hard, and it would be possible for an attacker to emulate a set of access points to transmit beacons with proper $T_{AP}^B$, which can mask the alteration in the manipulated GNSS time. 
Our framework can weigh alternative sources or even rank the verification external time sources, based not only on their precision but also on the perceived level of trust, as specified in Eq. \ref{eq:gt} (Sec. \ref{sec:multiple_ext_ts}). 

The future version of the implemented system will consider the possibility of adopting more types of external time sources, along with varying levels of trustworthiness. As an example, Long Range (LoRA) \cite{LoraAlliance2017} or IEEE 802.15.4 compliant networks \cite{wpan2015ieee}, can provide low power consumption connectivity. These technologies are potential external time sources to enhance the verification capability of the system once the infrastructure is deployed and the technologies gets popular in mobile systems. 

\section*{Acknowledgments} 
Work supported by the Swedish Foundation for Strategic Research (SSF) SURPRISE project and the KAW Academy Fellowship Trustworthy IoT project.

\bibliographystyle{IEEEtran}
\bibliography{ref}

\end{document}